\documentclass[prl,twocolumn]{revtex4}

\usepackage[dvips]{graphicx}%
\usepackage{amsmath,amssymb}
\usepackage{bm}

\begin{document}
  \small
\title{Fidelity criterion for quantum-domain transmission and storage of coherent states beyond unit-gain constraint} 

\author{Ryo Namiki}
\email[Electric address: ]{namiki@qi.mp.es.osaka-u.ac.jp}

\author{Masato Koashi}

\author{Nobuyuki Imoto}
\affiliation{CREST Research Team for Photonic Quantum Information, Division of Materials Physics, 
Department of Materials Engineering Science, 
Graduate school of Engineering Science, Osaka University, 
Toyonaka, Osaka 560-8531, Japan}
\date{August 11, 2008}
\begin{abstract} 
We generalize the experimental success criterion for quantum teleportation/memory in continuous-variable quantum systems to be suitable for non-unit-gain condition by considering attenuation/amplification of the coherent-state amplitude. The new criterion can be used for a non-ideal quantum memory and long distance quantum communication as well as quantum devices with amplification process. It is also shown that the framework to measure the average fidelity is capable of detecting all Gaussian channels in quantum domain. 
  \end{abstract}

\pacs{03.67.Dd, 42.50.Lc} 
\maketitle



In quantum information science \cite{NC00}, the manipulation of a quantum system is considered to be a channel that transforms a set of quantum states to another set of quantum states.
A fundamental distinction is posed on the channel whether it can be simulated by a \textit{measure-and-prepare} (M\&P) scheme or not. The M\&P scheme implies that the output of the channel is produced merely based on the classical data processing from the measurement outcomes and that the channel action breaks quantum entanglement shared between the input of the channel and other systems \cite{16}. Therefore, a natural benchmark for the \textit{quantum-domain} (QD) operation of a given experimental quantum manipulation is that the channel is outperforming any M\&P scheme. 
The M\&P scheme is an intercept-resend attack in the context of the quantum key distribution (QKD), and no secret key can be generated if the input-output relation is explained by an M\&P scheme. Hence confirming the QD operation is an important prerequisite for any QKD \cite{Cur04,Rig06}. Hammerer \textit{et al.} \cite{Ham05} have established the criterion for the QD operation of continuous-variable (CV) channels by proving a limit of the average fidelity achievable by the M\&P schemes, $F_c$, assuming an input set of coherent states: Surpassing $F_c$ ensures the QD operation for transmission and storage of coherent states. This criterion gave a proof for the long-standing conjecture of CV quantum teleportation about $F_c$ \cite{Bra00,Furusawa98} and provided a firm foundation for the central claims of experimental CV quantum teleportation \cite{Furusawa98,Zhan03,Bowen03e} and quantum memory \cite{Jul04}.  However, the application of this criterion is limited to the \textit{unit-gain} (UG) channels where the coherent-state amplitudes of the input state and the output state  are expected to be the same \cite{Bowen03,Zhan03}.

Quantum memory for light (QM) is a challenging protocol \cite{Koz00,Fle02}, which requires UG operation and involves not only storage of the states but also transfer of the states between different physical media, such as an optical system and an atomic ensemble. 
In any implementation \cite{Jul04,Liu01,Alex06,Kraus06}, it is likely that the effect of linear loss or damping of coherent-state amplitude becomes more significant as the storage period becomes longer. Therefore, some mechanism of gain adjustments seems to be necessary for the complete demonstration of QM. With gain control of the teleportation-based state transfer \cite{Koz00}, an above-$F_c$ operation of QM has been reported \cite{Jul04}.
Another possible solution for the gain control is to employ an amplifier \cite{amp}, where it is shown that the quantum-limit phase-insensitive amplifier can be implemented by linear optics and homodyne detection \cite{amp2}. 
However, the loss of signal information cannot be recovered by such gain-control scheme.  Rather, the conventional gain-control process imposes extra noise and acts so as to decrease the quantum correlations since the process is a non-unitary local operation. 
For example, coherent states sent through a lossy channel
with the transmission $\frac{1}{2}$ are still useful for quantum protocols because it is well-known that CV QKD with coherent states works in the presence of more-than-3dB loss \cite{hirano03,postsel}. But if we amplify the output states by a factor of 2 to establish a UG channel as a whole, the same final states can be obtained by a M\&P scheme where the classical outcome of a double homodyne measurement is sent through a classical communication channel \cite{Bra00}. This implies that the amplification process unnecessarily destroys the QD performance of a lossy channel.  
   Hence, the essential utility of the channels cannot be improved by the gain control and should be specified apart from the UG condition. 
  In practical side, it is not always easy to design a gain-control scheme compatible with the given process of storage, and further extra noise is imposed by experimental imperfections without saying its cost of experimental control.   
In addition, most of communication protocols are designed to be tolerant against losses in the optical fibers, and moderate losses in QMs used in such protocols should be acceptable as well.  Hence, it is not essential to focus on the UG channel, but non-UG devices should be investigated as well.
 There are several efforts to ensure ``quantum-regime'' operation of the storage of light, particularly in the electromagnetically induced transparency (EIT) approaches \cite{Liu01,Aka04,Hsu06}. For instance, storage of the squeezed state was experimentally demonstrated \cite{Aka04}. Moreover, there is a report of an EIT experiment with an estimation based on a different ``teleportation criteria'' \cite{Hsu06}. However, these efforts need not ensure the QD operation. 
Therefore, it is widely useful to present an experimental criterion that ensures the QD operation of the non-UG quantum devices. These motivate us to generalize the fidelity criterion to be suitable for the non-UG condition, especially in the presence of loss.

In this Letter, we provide a general classical boundary of the fidelity by considering attenuation/amplification of the coherent-state amplitude. The new criterion is essentially capable of detecting all the Gaussian channels in quantum domain, and is useful for experiments of transfer, storage, and transmission with dissipation.

 We define the average fidelity for the transformation task from a set of input states $\{ | \psi_x \rangle \}$ to a set of ideal output states $\{| \psi_x' \rangle \}$, 
\begin{eqnarray}
\bar F &=& \sum _x p_x \langle \psi_x'| E \left( |\psi_x \rangle\langle \psi_x |  \right) | \psi_x'\rangle  
\end{eqnarray} where $p_x$ is the prior distribution of the input states and $E$ is the channel action described by a completely positive trace-preserving (CPTP) map \cite{NC00}.  Whereas $E$ is a physical map, the task $\{ | \psi_x  \rangle \}  \to \{ | \psi_x' \rangle \} $ can be a physically impossible (unphysical) map, such as perfect cloning for a set of non-orthogonal states $\{ | \psi_x \rangle \}  \to \{ | \psi_x \rangle ^{\otimes 2} \}  $. The channel is simulated by the M\&P scheme when we can write \begin{eqnarray}
 E \left( |\psi_x \rangle\langle \psi_x |  \right)  &=& \sum _k  \langle \psi_x | \hat M _k| \psi_x  \rangle
\hat \rho_k \end{eqnarray} where $\{ \hat M_k \}$ is a positive-operator valued measure (POVM) and $\hat \rho_k $ is a density operator.  
The classical boundary of the average fidelity for the task $\{ | \psi_x  \rangle \} \to \{| \psi_x' \rangle \} $ is defined by the optimization over the M\&P schemes: 
\begin{eqnarray}
F_{c} 
 &\equiv & \sup_{\{\hat M_k  \},\{\hat \rho_k \} }\sum_{x,k} p_x  \langle \psi_x  | \hat M_k |\psi_x \rangle     \langle\psi_x'  |\hat \rho_k |\psi_x' \rangle   \label{classf} . \end{eqnarray}

We consider a transformation task of amplitude modulation  by the factor of $\eta > 0 $ for the coherent states:  $ \{ | \alpha \rangle  \} \to  \{ | \sqrt \eta \alpha \rangle \} $ assuming the prior distribution of a symmetric Gaussian function \begin{eqnarray}
p( \alpha ) =  \frac{\lambda }{\pi} \exp (- \lambda |\alpha |^2 ),\end{eqnarray} which represents the uniform distribution of coherent states in the limit $\lambda \to 0$. The task can be achieved by a lossy channel when $0 < \eta \le 1 $, while it becomes an unphysical noiseless amplification of coherent-state amplitude and is never achieved faithfully by any CPTP when $\eta > 1 $. 
In the following, using the method of \cite{Ham05}, we show that the classical boundary $F_{ \eta, \lambda}$ of the fidelity for this task is given by\begin{eqnarray}
F_{ \eta, \lambda}  
 &\equiv & \sup_{\{\hat M_k  \},\{\hat \rho_k \} }\sum_k\int d^2 \alpha p( \alpha )   \langle  \alpha |  \hat M_k |  \alpha \rangle     \langle \sqrt \eta \alpha  |\hat \rho_k | \sqrt \eta \alpha \rangle . \nonumber\\ &=& F(  \eta , \lambda) \equiv \frac{1+ \lambda}{1+ \lambda + \eta }.   \label{o2}\end{eqnarray}%
%
 
 \textit{Proof.} We decompose $ \hat M_k $ and $\hat \rho_k $ of $F_{ \eta, \lambda} $ by rank-1 projections as $ \hat M_k  = \sum_l |r_{k_l}\rangle \langle r_{k_l} |  $ and $\hat \rho_k = \sum_{j} p_{k_j} |\chi _{k_j}\rangle \langle\chi _{k_j}  | $ with $\sum_j p_{k_j} =1 $ and $\langle\chi _{k_j}|\chi _{k_j}\rangle =1 $, and define $| \phi_y \rangle \equiv \sqrt {p_{k_j}} |r_{k_l}\rangle $ and $|\chi _y\rangle \equiv |\chi _{k_j}\rangle $. The condition of POVM $\sum_k  \hat M_k  =\openone $ yields $  \sum_{y}|\phi_y\rangle\langle \phi_y| =\openone$. Then, $F_{ \eta ,\lambda } $ is expressed by a simpler form 
\begin{eqnarray}
F_{ \eta , \lambda } 
= \sup_{\left\{ \{ | \phi_y \rangle \}| \sum_{y}|\phi_y\rangle\langle \phi_y| =\openone \right\}} \sum_y ||\hat A_{\phi_y} ||_\infty \label{eq6}\end{eqnarray}
where %
$\hat A_{\phi_y} \equiv \int d \alpha p( \alpha )  |\langle  \alpha  |\phi_y\rangle |^2   | \sqrt \eta \alpha \rangle \langle  \sqrt \eta \alpha | \ge 0$ 
 and  $p$ norm of an operator $\hat A $ is defined by
$||\hat A  ||_p \equiv [ \textrm{Tr}(| \hat A |^p)]^{1/p}$. 
Note that the optimization over $|\chi_y \rangle $ is absorbed in the operator norm. 

In order to estimate the norm we introduce the operator $\hat B$ that satisfies
\begin{eqnarray}
\left( ||\hat A_{\phi } ||_p \right) ^p = \textrm{Tr}(|\phi \rangle \langle \phi |^{\otimes p}    \hat B  ) \label{pno2}. 
\end{eqnarray}
The operator $\hat B$ is explicitly written as
\begin{eqnarray} \hat B &\equiv&  \int d^2 \alpha_1 d^2 \alpha_2 \cdots d^2 \alpha_p \nonumber  \langle  \sqrt \eta \alpha_1 |  \sqrt \eta  \alpha_2 \rangle  \langle  \sqrt \eta  \alpha_2 |  \sqrt \eta  \alpha_3 \rangle\\  & &\times \cdots \langle  \sqrt \eta  \alpha_p | \sqrt \eta  \alpha_1  \rangle p(\alpha_1) p(\alpha_2) \cdots  p(\alpha_p)  \bigotimes_{j=1}^{p}    |   \alpha_j\rangle \langle   \alpha_j |  \nonumber  \\ &=  &\int d^2 \vec{\alpha} \left({\frac{\lambda}{\pi  }}\right)^{p } \exp\left( -  \vec{\alpha} ^\dagger M_{\lambda+\eta, \eta}  \vec{\alpha}  \right) \bigotimes_{j=1}^{p}    |\alpha_j\rangle \langle \alpha_j | \nonumber \end{eqnarray}
where we wrote $\vec{\alpha} = (\alpha_1 ,\alpha_2 , \cdots , \alpha_p )^t$ and defined a $p \times p$ matrix 
$M_{\lambda, \eta }  \equiv  \lambda E _p - {\eta }  C$   
with the identity $E_p$ and the basic circulant permutation matrix $C$ whose element is $(C)_{k,l} = \delta_{k+1,l}+ \delta_{k+1-p,l}$. 
We can diagonalize $M_{\lambda+\eta ,\eta }$ by a $p \times p$ unitary matrix $V$ whose element is $(V)_{k,l}=e^{-2\pi i kl/p} /\sqrt p $. The $j$-th eigenvalue of  $M_{\lambda+\eta ,\eta }$  is 
$\chi_j  \equiv  \lambda   + \eta  ( 1 +e^{ {2\pi i  j}/{p} } )$. 
We introduce a new basis of modes defined through their
annihilation operators $\vec b = (\hat b_1 ,\hat b_ 2 , \cdots , \hat b_p )^t = V \vec a $, 
 where $\vec{a} = (\hat a_1 ,\hat a_ 2 , \cdots , \hat a_p )^t$
is the annihilation operators of the original set of modes.
Then the $p$-mode coherent state $\bigotimes_{j=1}^{p}    |\alpha_j\rangle \langle \alpha_j | $ in basis $\vec{a}$ is also expressed as
a $p$-mode coherent state $\bigotimes_{j=1}^{p}    |\beta_j\rangle \langle \beta_j |$ in basis $\vec{b}$, where $ \beta_i \equiv \sum_{j=1}^p(V)_{i,j} \alpha_j $. Hence we find a decoupled expression of $\hat B =    \bigotimes_{j=1}^{p} {({\lambda}/{\pi   })} \int d^2 \beta_j     \exp ( -  {\chi_j}   | \beta_j | ^2  )    |\beta _j\rangle \langle \beta_j | $.
Expanding $|\beta_j \rangle$ by Fock-basis %
and performing the integration of $\beta$ noting that $\Re e \left( 1+ \chi_j  \right) >0 $,
we obtain the diagonal expression
\begin{eqnarray}
 \hat B 
&= &   \bigotimes_{j=1}^{p}      {\frac{\lambda    }{ 1 + \chi_j   }} \sum_{n_j=0}^{\infty} \left( \frac{1}{1+  \chi_j   } \right)^{n_j}    |n _j\rangle \langle n_j |  \nonumber \\ 
&= &  {\frac{\lambda ^p  }{(1+ \lambda   +\eta)^p - \eta ^p}} \bigotimes_{j=1}^{p}   \sum_{n_j=0}^{\infty} \left( \frac{1}{ 1+  \chi_j    } \right)^{n_j}    |n _j\rangle \langle n_j |  \nonumber 
\end{eqnarray} where $  |n _j\rangle  $ represents Fock states of the new $j$-th mode and we used the relations $\det (M_{\lambda, \eta}) =\lambda ^p - \eta ^p $ and $\prod_{j=1}^{p} (1 +\chi_j) = \det (M_{1 + \lambda + \eta , \eta} )$ to determine the factor before the product. 
Since $ 1 + \lambda   \le | 1 + \chi_j | $, we find 
\begin{eqnarray}
 | \hat B |
& \le &  {\frac{\lambda ^p  }{(1+ \lambda   +\eta)^p - \eta ^p}} \bigotimes_{j=1}^{p}   \sum_{n_j=0}^{\infty} \left( \frac{1}{ 1+  \lambda    } \right)^{n_j}    |n _j\rangle \langle n_j |  \nonumber \\ 
&= &   {\frac{(1+ \lambda ) ^p  }{(1 + \lambda +\eta)^p - \eta ^p}}       \bigotimes_{j=1}^p \hat \rho_ \lambda    \label{bigb1}\end{eqnarray}
where 
we introduced normalized thermal states 
$ \hat \rho_\lambda  \equiv  { [{\lambda    }/{( 1+ \lambda)]   }} \sum_{n =0}^{\infty} ({ 1+ \lambda  } )^{-n }    |n  \rangle \langle n  |   .$
  Since $|\phi\rangle\langle \phi |^{\otimes p}$ is positive and $\hat B$ is normal, we can verify 
  $\textrm{Tr}  (|\phi\rangle\langle \phi |^{\otimes p} \hat B  )  \le  \textrm{Tr} (|\phi\rangle\langle \phi |^{\otimes p} | \hat B | )$. 
From Eqs. (\ref{pno2}), (\ref{bigb1}) and this inequality, we obtain
\begin{eqnarray}
\left( || \hat A_\phi ||_p \right)^p   
&\le&    {\frac{(1+ \lambda ) ^p  }{(1 + \lambda +\eta)^p - \eta ^p}}     \textrm{Tr} \left\{ |\phi\rangle\langle \phi |^{\otimes p}    \bigotimes_{j=1}^p \hat \rho_ \lambda   \right\} \nonumber \\
&=&   
 {\frac{(1+ \lambda ) ^p  }{(1 + \lambda +\eta)^p - \eta ^p}}     \left( \textrm{Tr} \left\{ |\phi\rangle\langle \phi |   \hat \rho_ \lambda   \right\}  \right) ^p  \label{bp} 
\end{eqnarray} where
the final expression comes from the fact that the product of the thermal states in the modes $\vec b $ is also the product of the thermal states in the modes $\vec a $ with the same temperature. Inequality (\ref{bp}) implies
$ || \hat A_\phi ||_\infty  
 \le   F(  \eta , \lambda)    \textrm{Tr} \left\{ |\phi\rangle\langle \phi |   \hat \rho_ \lambda   \right\}$. 
 Using the relation $ \sum _y      |\phi_y\rangle\langle \phi_y |     = \openone $, we can bound the fidelity in Eq. (\ref{eq6}) by
$ F_{ \eta , \lambda } \le F( \eta, \lambda )  $. 

Next we show that we can achieve this bound by an M\&P scheme, which is the projection onto coherent states and preparation of states according to the outcome. The measurement operator can be represented by $\hat A_\alpha \equiv |\frac{\sqrt{\eta  } }{1 +\lambda  } \alpha \rangle \langle \alpha | /\sqrt\pi $. It gives the POVM element $\hat M_\alpha \equiv \hat A_\alpha^\dagger \hat A_\alpha = |\alpha \rangle \langle \alpha | / \pi   $
 and the corresponding state preparation $\hat \rho_\alpha \equiv  \hat A_\alpha \hat \rho \hat A_\alpha ^\dagger /\textrm{Tr} ( \hat A_\alpha \hat \rho  \hat A_\alpha ^\dagger )= |\frac{\sqrt{\eta  } }{1 +\lambda  } \alpha \rangle \langle \frac{\sqrt{\eta } }{1 +\lambda  } \alpha |  $  where $\hat \rho $ is an arbitrary density operator with $\textrm{Tr} ( \hat A_\alpha \hat \rho  \hat A_\alpha ^\dagger ) > 0$. This M\&P scheme corresponds to the intercept-resend attack proposed in \cite{namiki2} when $\lambda =0$. The fidelity of this M\&P scheme is 
straightforwardly calculated by using $|\langle  \alpha | \beta  \rangle|^2 =e^{-|\alpha -\beta | ^2} $, leading to 
$F_{ \eta, \lambda }      \ge     
\int\int p( \alpha ) |\langle \sqrt \eta \alpha| \hat A_\beta   |  \alpha \rangle   |^2    d^2 \beta   d^2 \alpha  = F(  \eta , \lambda)$. 
Therefore, we can conclude that $F_{  \eta, \lambda }   =F( \eta , \lambda)$. \hfill$\blacksquare$

An interesting application is the experimental success criterion for non-UG quantum devices, which include transfer, storage, transmission, amplification, and so on. Suppose that the input and the output of quadrature operators are related as $\hat X_{\textrm{out}} = \sqrt \eta \hat X_{\textrm{in}} + \hat X_{\textrm{noise}} $ by a positive gain factor $\eta$ and noise operator $\hat X_{\textrm{noise}} $.
 If the average fidelity $\bar F$ of the task $\{|\alpha \rangle\} \to \{|\sqrt \eta \alpha \rangle \}$ surpasses the classical boundary $F ( \eta , \lambda )$, it ensures that no M\&P scheme can simulate the experiment.
Note that the classical bound on the fidelity for a more general class of tasks $\{ \hat U |  \alpha \rangle \} \to \{ \hat V | \sqrt \eta  \alpha  \rangle \} $ is also $F( \eta, \lambda )$ where $\hat U $ and $\hat V$ are any unitary operators. This is because the replacement $\{ \hat M_k , \hat \rho_k \} \to \{\hat U ^\dagger \hat M_k \hat U ,  \hat V ^\dagger \hat \rho_k  \hat V \}$ provides the same optimization problem. Thus, we can select $\hat U$ and $\hat V$ as well as $\eta $ to ensure the QD   operation. 

 This framework is general enough to detect all Gaussian channels in the quantum domain, which is proved as follows.
Let us write the quadrature operators
$\hat{x}_{+}\equiv (\hat{a}+\hat{a}^\dagger)/\sqrt{2}$
and $\hat{x}_{-}\equiv (\hat{a}-\hat{a}^\dagger)/\sqrt{2}i$
as a column vector
$\hat R\equiv (\hat{x}_+ ,\hat{x}_-)^t$.
The characteristic function of a density operator $\hat\rho$
is defined by $\phi(z)\equiv \textrm{Tr}[\hat\rho \exp(i\hat R^t z)]$,
where $z$ is a real column vector. Then, $\hat\rho$ is
a Gaussian state when it has the form
$\phi(z)=\exp(id^tz-\frac{1}{2}z^t\gamma z)$.
Here $d=\textrm{Tr}(\hat\rho \hat R)$ is the mean amplitude, and
$\gamma = \textrm{Tr}\hat\rho [{\hat R \hat R^t+(\hat R \hat R^t)^t} ]/{2} - d d^t$ 
is the correlation matrix \cite{Gchannel1,Holevo1}.
The physical requirement of the uncertainty relation is
given by the condition $\gamma\ge (i/2) \Delta$ with
$ \Delta\equiv  \left(\begin{array}{cc}    0&-1  \\ 1&0 \end{array}\right)$.
%
The Gaussian channel is a CPTP map that transforms the Gaussian states into the Gaussian states \cite{Gchannel1,Holevo1}. 
   Except for the freedom of the uniform displacements, any one-mode Gaussian channel
can be described \cite{Holevo1} by a pair of 2$\times$2 matrices $(K,M)$ that transforms $\gamma $ and $d$ as $\gamma' = K^t \gamma K +  M$ and $d' = K d$, respectively. In order that $\gamma'$ is physical for any physical $\gamma$, $M\ge (i/2)(\Delta-K^t\Delta K)$ must be satisfied.  
 Recently it is shown \cite{Holevo2} that, with appropriate choice of  $\hat U$ and $\hat V$, the pair $(K,M)$ of any one-mode QD Gaussian channel becomes either (i) $K=E_2$,  
$M  = \left(\begin{array}{cc  }     1/2  & 0    \\     0  & 0  \end{array}\right)$ or (ii) $K =  \sqrt{\eta} E_2$, $M=( \tilde n+|1-\eta |/2) E_2$ where $E_2$ is the identity matrix and $0 \le\tilde n < \min\{1,\eta \}$.   
The channel (i) adds a constant Gaussian noise onto a single quadrature component. The channel (ii) is Gaussian amplification/attenuation with the gain $\eta$. In the limit $\lambda \to 0 $ both cases are detectable by our criterion: By choosing the task $\{|\alpha\rangle\} \to \{|\alpha\rangle\}$, the fidelity of the channel (i) is $\bar F = \left[ \det (\gamma + \gamma') \right]^{- 1/ 2 } =  { ({2}/{3})^{1/2}}$ where  $\gamma = E_2 /2$ is the correlation matrix of the coherent-state input. The fidelity is strictly higher than the classical boundary of $F(1,0) = 1/2$; By choosing the task $\{|\alpha\rangle\} \to \{|\sqrt{\eta} \alpha\rangle\}$, the fidelity of the channel (ii) is $\bar F=  \left[ \det (\gamma + \gamma') \right]^{- 1/ 2 }  = 2/( 1+\eta +|1-\eta|+2 \tilde n )  $, and is higher than the classical boundary of $F(\eta, 0)= {1}/({1+\eta})$ if $\tilde n < \min\{1,\eta \}$.

In experiments, fidelity to a coherent state $|\alpha \rangle$ is directly determined by measuring the probability of photon detection after the displacement operation $\hat D(- \alpha ) $ that displaces $| \alpha \rangle$ to $|0\rangle $. In CV systems, homodyne measurement is commonly used and it might be useful to provide similar criterion in terms of the quadrature variances. %
Let us define the quadrature mean-square deviation from $\sqrt \eta \alpha $ by $V_\pm \equiv \langle \Delta \hat x_\pm^2 \rangle +  (\langle \hat x_\pm \rangle  - \sqrt \eta \alpha _ \pm   )^2$ 
 associated with the input $|\alpha \rangle $ and output $\hat \rho _\alpha \equiv E(|\alpha \rangle \langle \alpha |)$, where $ \langle \cdot  \rangle \equiv \textrm{Tr}(\cdot \hat \rho_\alpha ) $ 
  and $\alpha_\pm \equiv \langle \alpha | \hat x_\pm  | \alpha \rangle $. %
  Since the sum of $V_\pm $ is $ v \equiv  V_ + + V_ -     = \textrm{Tr} [(2 \hat a ^\dagger  \hat a +1  )  \hat D (-\sqrt \eta \alpha  ) \hat \rho_\alpha \hat D^\dagger (-\sqrt \eta \alpha  ) ] $ and  $\textrm{Tr}(\hat a^\dagger \hat a \hat\rho) \ge 1 - \textrm{Tr}(|0 \rangle \langle 0 | \hat \rho)$ for any state $\hat \rho $, we can bound $ v   \ge    3 -2    \langle \sqrt \eta \alpha  | \hat \rho_\alpha   | \sqrt \eta \alpha  \rangle$. Averaging both sides over $\alpha$ with $p(\alpha ) $, we find $\bar \delta \equiv \bar v   -1  \ge 2 (1 -\bar F)$. Then we can see that $\bar F  > F( \eta, \lambda)$ is satisfied if 
\begin{eqnarray} \bar \delta < 2 (1- F(  \eta, \lambda) ) = \frac{2 \eta }{ 1+ \lambda + \eta}.  \end{eqnarray} 
This is a sufficient condition for the QD operation of the physical map $E$.

If we consider a more general task $ \{ | \sqrt N \alpha \rangle  \} \to  \{ | \sqrt \eta \alpha \rangle \} $ for $N>0$, the classical boundary is shown to be $F(\eta/N, \lambda /N)$ since the definition is obtained by  the replacement $(\alpha, \eta, \lambda) \to (\sqrt N \alpha, \eta /N, \lambda /N )$ at Eq. (\ref{o2}). For $\eta =1 $ and assuming that $N$ is a positive integer, since the transformation from $|\sqrt N \alpha \rangle$ to $|\alpha \rangle^{\otimes N} $ is reversible, the fidelity $F(1/N, \lambda /N)$ corresponds to the optimal fidelity for the coherent-state estimation from $N$ copies with the prior distribution $p(\alpha )$ \cite{Nis06}. $F(1/N, \lambda /N)$ is also the optimal fidelity of the $N$-to-$\infty$ cloning of coherent states with the prior distribution $p(\alpha )$ since the state estimation and the asymptotic cloning are equivalent for any given ensemble of pure states \cite{Bae06}. Since we have derived $F(1/N, \lambda /N)$ without any assumption on the M\&P scheme, neither Gaussian nor non-Gaussian cloning machine can surpass $F(1/N, \lambda /N)$. On the other hand, the optimal M\&P scheme, which is given by the measurement operator $\hat A_ \alpha  '\equiv |\frac{\sqrt{N} }{N + \lambda}  \alpha \rangle \langle \alpha | /\sqrt \pi $, is a Gaussian operation. Therefore, in the $N$-to-$\infty$ cloning of coherent states with the prior distribution $p(\alpha )$, non-Gaussian cloner \cite{Cer05} does not outperform Gaussian cloner, and for the flat distribution $\lambda \to 0$, the Gaussian cloning machine proposed in \cite{Cerf00,optc1} is optimal.

 While the above discussions concern the optimization over the M\&P schemes and to derive a classical boundary on fidelity, one may also ask what is the quantum  bound on the fidelity for an unphysical task, namely, the fidelity optimized over the whole CPTP maps. For the amplification task $\{| \alpha\rangle \}\to \{|\sqrt \eta \alpha\rangle  \} $ with $\eta>1$, the problem is equivalent to the optimal cloning of coherent states with respect to the joint fidelity \cite{Cer05} when $\lambda \to 0$ and $\eta$ is an integer $N' > 1$ due to the reversibility $|\sqrt{ N'} \alpha \rangle \leftrightarrow  |\alpha\rangle^{\otimes N'}$. Since there is no reason to restrict $N'$ to be an integer in the proof \cite{Cer05} for the joint fidelity case, we can say that the quantum optimal amplifier is given by a Gaussian amplifier, which is the one in the case (ii) with $\tilde n=0$ above, achieving the optimal value of $ \bar F =  1/\eta $.  The Gaussian amplifier is also shown to be optimal when the figure of the merit is quadrature noise \cite{amp} or trace distance \cite{Gut06}. The analysis for the finite distribution $\lambda >0$ is in progress.

 In conclusion, we have provided a general boundary of the average fidelity achieved by the M\&P schemes assuming a transformation task that modulates coherent-state amplitude. The formula can be applicable to the experimental success criterion for continuous-variable quantum devices with dissipation/amplification, such as storage of light for a quantum memory. We have also shown that the framework to measure the average fidelity for the transformation task is capable of detecting all Gaussian channels in quantum domain.

We thank Mikio Kozuma for helpful discussions. This work was supported by a MEXT Grant-in-Aid for Young Scientists (B) 17740265.

\end{document}